\documentclass[journal]{IEEEtran}

\usepackage{xcolor,soul,framed} 

\colorlet{shadecolor}{yellow}
\usepackage[pdftex]{graphicx}
\graphicspath{{../pdf/}{../jpeg/}}
\DeclareGraphicsExtensions{.pdf,.jpeg,.png}

\usepackage[cmex10]{amsmath}
\usepackage{array}
\usepackage{mdwmath}
\usepackage{mdwtab}
\usepackage{eqparbox}
\usepackage{url}
\usepackage{xcolor}
\usepackage{soul}
\usepackage{subfigure}
\usepackage{graphicx}
\usepackage{float}

\begin{document}
\bstctlcite{IEEEexample:BSTcontrol}
    \title{Perfect anomalous reflection using compound metallic metagratings}
  \author{Mahdi~Rahmanzadeh,
      and~Amin~Khavasi

    \thanks{The authors are with the Department of Electrical Engineering, Sharif University of Technology, Tehran 11155-4363, Iran (email:rahmanzadeh.mahdi@ee.sharif.edu ; khavasi@sharif.edu}}

\maketitle

\begin{abstract}
Metagrating is a new concept for wavefront manipulation that, unlike phase gradient metasurfaces, does not suffer from low efficiency and also has a less complicated fabrication process. In this paper, a compound metallic grating (a periodic structure with more than one slit in each period) is proposed for anomalous reflection. The electromagnetic response of this grating is analyzed by a proposed analytical method and then a closed-form and analytical expressions are presented for the reflection coefficients of the higher diffracted orders. The proposed method is verified by full-wave simulations and the results are in excellent agreement. Thanks to the geometrical asymmetry of compound metallic grating, we use it for designing anomalous reflection in the normal incident. Given analytical expressions for reflection coefficients, a perfect anomalous reflector is designed via transferring all the incident power to $(-1)$ diffraction order. The structure designed in this study has an unprecedented near-to-unitary efficiency of $99.9\%$. Finally, a multi-element compound metallic grating is proposed for reflecting the normal incident to angles of below $30^\circ$, which is a challenging accomplishment. This excellent performance of compound metallic grating shows its high potential for microwave and terahertz wave-front manipulation applications. 
\end{abstract}

\section{Introduction}

\IEEEPARstart{T}{HE} anomalous reflection/refraction is an interesting subset of wave manipulations owing to its important role in laser, optics, and so on \cite{rose2005wavelength,hawthorn2001littrow,rhee1994chirped}. For a long time, this attractive phenomenon has been realized by means of reflected arrays \cite{hum2007modeling}. Nowadays, due to the development of nanotechnology, metasurfaces are common elements to achieve anomalous reflection \cite{li2015visible,sun2012high,wong2018perfect}. 

Metasurfaces, two-dimensional artificial structures used for manipulating the propagation of electromagnetic (EM) waves, have attracted much attention because of their numerous applications\cite{li2018new,mueller2017metasurface,rahmanzadeh2018multilayer,rahmanzadeh2018adopting,momeni2019generalized,momeni2018information,kildishev2013planar,sautter2015active,zheng2015metasurface,hosseininejad2019digital}. More specifically, phase gradient metasurface (PGM) can provide exceptional phenomena such as surface wave conversion \cite{sun2012gradient}, vortex beam generation \cite{rouhi2019multi}, and beam focusing \cite{li2015x}. Also, PGM can realize the anomalous reflection based on generalized Snell's law, which is achieved by the inhomogeneous phase profile of metasurfaces \cite{yu2011light}. Highly efficient anomalous reflection is a strict requirement for the above-mentioned applications. However, metasurfaces cannot achieve highly efficient until they are active and extremely local \cite{asadchy2016perfect,epstein2016huygens,diaz2017generalized}. Moreover, the practical implementation of metasurfaces has significant difficulties. 

 Recently Ra'di \textit{et al} introduced the concept of metagrating, which enables manipulations of EM waves with unitary efficiency \cite{ra2017metagratings}. In contrast to metasurfaces, metagrating does not need deeply subwavelength elements, making their fabrication process less complicated. The concept of metagrating is similar to the diffraction grating which has been one of the attractive research topics in the past decades \cite{wood1902remarkable,khavasi2009adaptive,harvey1991external,blanchard1999simultaneous}. The design principles of metagratings are related to Floquet-Bloch (FB) theory. Based on FB theory, when a plane wave impinges on a periodic structure, a discrete set of diffracted waves can be generated that some of them are propagating and others are evanescent. The number of propagating and evanescent waves is determined by the period of the structure and angle of incidence. To achieve anomalous reflection, all propagating modes should be vanished (including the specular mode) except one of them and all the power is transferred to that propagating non-specular FB mode.

 Inspired by this idea, several anomalous reflectors have been proposed but most of them are designed through numerical simulations \cite{ra2018reconfigurable,dong2019low,sell2018ultra,neder2019combined,inampudi2018neural}. Although some analytical methods are proposed, all of them have some problems that restrict their performance. For example, \cite{chalabi2017efficient} considered a metagrating formed by pairs of isotropic dielectric rods that suspended in free space above a perfect electric conductor (PEC) ground, which is not practical for implementation. Furthermore, the analytical method presented in \cite{rabinovich2018analytical,rabinovich2019arbitrary,rabinovich2019experimental}, due to its geometric symmetry, cannot be used in the case of  normal incident, which is the most important case in the anomalous reflectors. After all, the power efficiency of the previously proposed metagrating-based anomalous reflectors is limited to $94\%$\cite{rabinovich2018analytical,ra2018reconfigurable,chen2018polarization}. Moreover, none of the proposed structures provide anomalous reflection below $30^\circ$. This case is more challenging, since $(\pm2)$ FB modes become propagating. Although a conceptual design is proposed based on multiple electric line currents for this case\cite{popov2018controlling}, however a practical structure with high efficiency has not been presented to the best of the author's knowledge.

 In this paper, we show that achieving to unitary efficiency for anomalous reflection is possible using compound metallic gratings (CMGs). A CMG includes finite numbers of slits in each period, which can be used as a reflection grating (consists of slits carved out of a thick metal slab ) or transition grating (connects two separate open regions through groups of slits in a metal slab). In \cite{molero2016dynamical}, an analytical equivalent-circuit model has been proposed for compound  metallic gratings with an arbitrary number of slits per unit. However, higher diffracted orders cannot be calculated in this method. We extend this equivalent-circuit to include higher order diffracted modes in our method. Thanks to this analytical method, we design a metagrating with near-to-unitary efficiency ($99.9\%$) at normal incidence using a CMG. We also demonstrate anomalous reflection below $30^\circ$ with an impressive efficiency of $98.6\%$.

\section{An analytical method for analysis of CMGs}

\subsection{Scattering from compound metallic grating}

\begin{figure}
\centering \includegraphics[width=\linewidth]{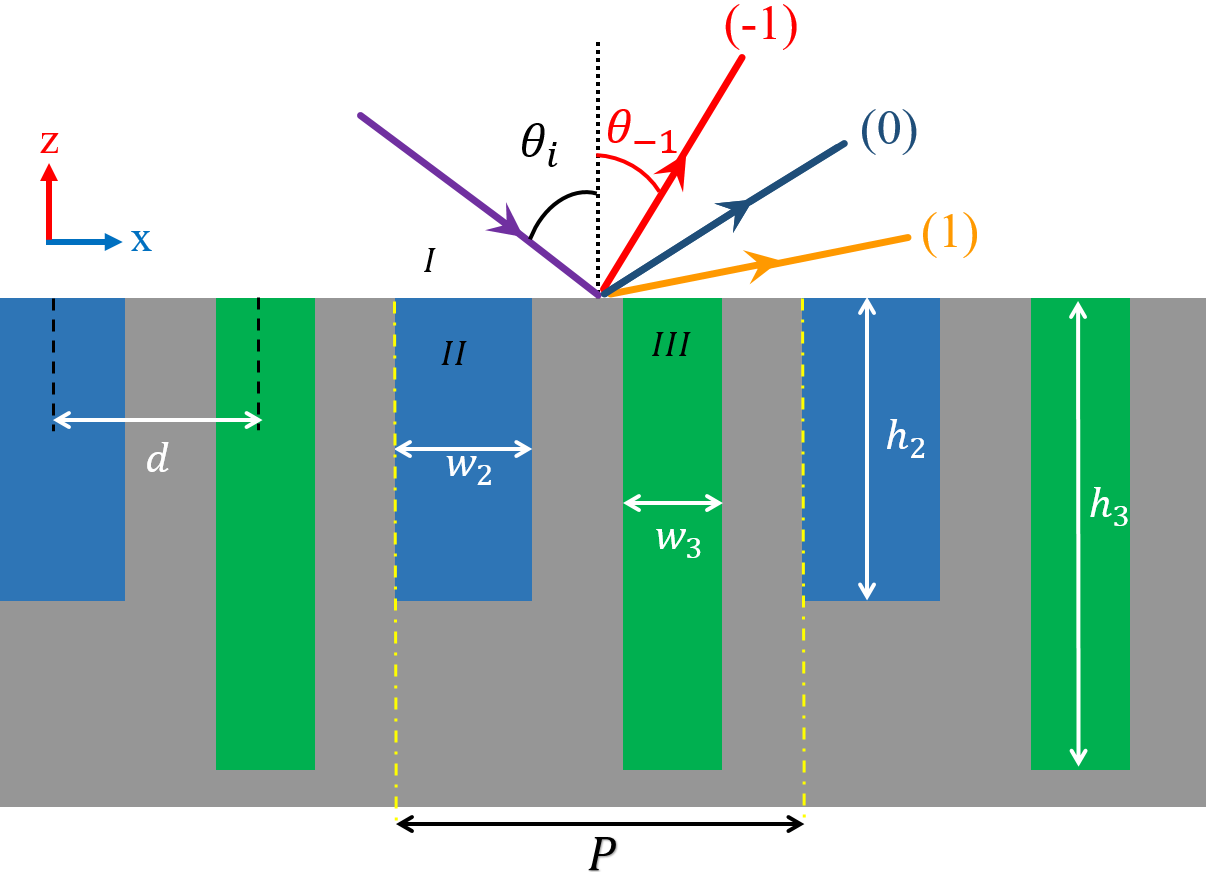}
\caption{The structure of CMG with two slits per period. CMG is illuminated by an oblique TM polarized wave  with angle $\theta_i$}
\label{CMG}
\end{figure}

The schematic of CMG, which includes two slits in a period, is illustrated in Fig.\ref{CMG}. Each unit cell consists of two slit regions with widths of $w_1$ and $w_2$ filled with a dielectric medium, the refractive indices of these regions are $n_2$ (region II) and $n_3$ (region III), respectively. The distance between the center of slits is $d$ and the height of the first and second slit is $h_2$ and $h_3$, respectively (Fig.\ref{CMG}). The period of the array is $P$ in the x-direction. Also, the structure is assumed to be infinite along the y-direction ($\partial/\partial y\sim 0$ ). It should also be noted that the time-harmonic of the form $\exp(j{\omega}t)$ is assumed throughout this paper. 

Consider that the structure is illuminated by an oblique transverse-magnetic (TM) polarized plane wave (the magnetic field  is along the $y$ direction) propagating along the z-direction with, whose incidence angle is $\theta_i$. Based on FB theorem, the total field in the region $z>0$ is expressed by the Rayleigh expansion \cite{yarmoghaddam2014circuit}

\begin{subequations}
\label{RayExp}
  \begin{equation}
  \label{Hy1}
 {H_{1y}} = H_{10}^ + {e^{j{k_{z10}}z}}{e^{ - j{k_{x10}}x}} + \sum\limits_n {H_{1n}^ - {e^{ - j{k_{z1n}}z}}{e^{ - j{k_{x1n}}x}}} 
  \end{equation}
  \begin{equation}
  \label{Ex1}
 {E_{1x}} =  - {\xi _{10}}H_{10}^ + {e^{j({k_{z10}}z-{k_{x10}}x})} + \sum\limits_n {{\xi _{1n}}H_{1n}^ - {e^{ - j({k_{z1n}}z+k_{x1n}x)}}}
  \end{equation}
\end{subequations}
where $k_{z1n}$ and $k_{x1n}$ are the wave-vector components in $z$ and $x$ directions in  region I, respectively, and they are given by

\begin{subequations}
\label{wave number}
  \begin{equation}
  \label{kx1m}
k_{x1n}=k_{x10}+\frac{2n\pi}{P} ~~~;~~~(n=0,\pm1, \pm 2, ...)
  \end{equation}
  \begin{equation}
  \label{kz1m}
 k_{z1n}=-jk_{0} \sqrt{(n_1 sin(\theta_i)+\frac{n\lambda}{P})^2-n_1^2};~(n=0,\pm1,\pm 2, ...)
  \end{equation}
\end{subequations}
where
\begin{equation}
  \label{kx10}
k_{x10}=k_0 n_1 sin(\theta_i)
  \end{equation}
where $k_{0}=\omega (\varepsilon_{0}\mu_{0})^{1/2}$ is the free space wave number and $\lambda$ is
the free space wavelength. It should be noted in the all above equations, the subscript $n$ corresponds to the order of the diffracted wave. Furthermore, in \eqref{Ex1}
\begin{equation}
  \label{Zita1n}
\xi_{1n}=\frac{k_{z1n}}{\omega \varepsilon_0 n^2_1}
  \end{equation}
is the TM-wave admittance of the $n$th diffracted order in region I.

In regions II and III, we assume slits are in single-mode regime and thus only transverse electromagnetic (TEM) mode is propagation in these regions. Hence, the magnetic and electric fields in regions I and II can be written as

\begin{subequations}
\label{RegionII}
  \begin{equation}
  \label{HfieldRegionII}
{H_{2y}} = H_{20}^ + {e^{j{\beta _2}z}} + H_{20}^ - {e^{ - j{\beta _2}z}}
  \end{equation}
  \begin{equation}
  \label{EfieldRegionII}
{E_{2x}} =  - {\xi _{20}}H_{20}^ + {e^{j{\beta _2}z}} + {\xi _{20}}H_{20}^ - {e^{ - j{\beta _2}z}}
\end{equation}
\end{subequations}
for $x\in[0,w_2]$,and
\begin{subequations}
\label{RegionIII}
  \begin{equation}
  \label{HfieldRegionIII}
{H_{3y}} = H_{30}^ + {e^{j{\beta _3}z}} + H_{30}^ - {e^{ - j{\beta _3}z}}
  \end{equation}
  \begin{equation}
  \label{EfieldRegionIII}
{E_{3x}} =  - {\xi _{30}}H_{30}^ + {e^{j{\beta _3}z}} + {\xi _{30}}H_{30}^ - {e^{ - j{\beta _3}z}}
\end{equation}
\end{subequations}
for $x\in[w_2/2+d-w_3/2,w_2/2+d+w_3/2]$.

\par
where $\beta_i=k_0 n_i~~(i=2,3)$ is the propagation constant of the TEM mode supported by a parallel plate waveguide. In addition, $\xi_{i0} $ is the wave admittance of the medium inside each slit that is obtained by
\begin{equation}
  \label{Zitai0}
\xi_{i0}=\frac{\beta_i}{\omega \varepsilon_0 n^2_i} ~~~ (i=2,3)
  \end{equation}

Since we consider that only the fundamental mode is dominant and higher-order modes are in cut-off, the validity of this method is limited to the frequency range $\lambda>2w_i n_i ~~(i=2,3)$ \cite{yarmoghaddam2014circuit}.

 Now, by applying the appropriate boundary conditions at the $z=0$ for the electric fields (the continuity of the tangential electric field at every point of the unit cell) using {\eqref{Ex1}},{\eqref{EfieldRegionII}} and {\eqref{EfieldRegionIII}} we have

\begin{subequations}
\label{BCforE}
  \begin{eqnarray}
  \label{BCforE_0 order}
H_{10}^ +  - H_{10}^ -  = \frac{{{\xi _{30}}}}{{{\xi _{10}}}}H_{30}^ + M_{3 + }^0 - \frac{{{\xi _{30}}}}{{{\xi _{10}}}}H_{30}^ - \,M_{3 + }^0  \nonumber\\ + \frac{{{\xi _{20}}}}{{{\xi _{10}}}}H_{20}^ + M_{2 + }^0\, - \frac{{{\xi _{20}}}}{{{\xi _{10}}}}H_{20}^ - \,M_{2 + }^0
 \end{eqnarray}

  \begin{eqnarray}
  \label{BCforE_m order}
H_{1n}^ -  =  - \frac{{{\xi _{20}}}}{{{\xi _{1n}}}}H_{20}^ + M_{2 + }^n + \frac{{{\xi _{20}}}}{{{\xi _{1n}}}}H_{20}^ - M_{2 + }^n \nonumber \\ - \frac{{{\xi _{30}}}}{{{\xi _{1n}}}}H_{30}^ + M_{3 + }^n + \frac{{{\xi _{30}}}}{{{\xi _{1n}}}}H_{30}^ - M_{3 + }^n ~ ~ ~ n \ne 0
\end{eqnarray}
\end{subequations}

where
\begin{subequations}
\label{M+-}
  \begin{equation}
  \label{M2+-}
M_{2 \pm }^n = \,\frac{1}{P}\int\limits_0^{{w_2}} {{e^{ \pm j{k_{x1n}}x}}dx} \,
  \end{equation}
  \begin{equation}
  \label{M3+1}
M_{3 \pm }^n = \frac{1}{P}\int\limits_{0.5{w_2} + d - 0.5{w_3}}^{0.5{w_2} + d + 0.5{w_3}} {{e^{ \pm j{k_{x1n}}x}}dx} 
\end{equation}
\end{subequations}
which are obtained by multiplying the electric fields to $e^{jk_{x1n}x}$ and taking the integral of both sides over one period. Using boundary conditions of a tangential magnetic field and {\eqref{Ex1}}, {\eqref{EfieldRegionII}}, {\eqref{EfieldRegionIII}} and then taking the integral of both sides over each slit width yields
\begin{subequations}
\label{BCforH}
  \begin{equation}
  \label{BCforH_2}
P\,H_{10}^ + M_{2 - }^0 + P\,\sum\limits_n {H_{1n}^ - M_{2 - }^n}  = {w_2}H_{20}^ +  + {w_2}H_{20}^ -
  \end{equation}
  \begin{equation}
  \label{BCforH_3}
P\,\,H_{10}^ + M_{3 - }^0 + P\sum\limits_n {H_{1n}^ - M_{3 - }^n}  = {w_3}H_{30}^ +  + {w_3}H_{30}^ -
\end{equation}
\end{subequations}
Finally, we apply perfect electrical conductor (PEC) boundary conditions at the end of each slit, which leads to
\begin{subequations}
\label{PEC_bc}
  \begin{equation}
  \label{PEC_bc_2}
H_{20}^ -  = H_{20}^ + {e^{ - 2j{\beta _2}{h_2}}}
  \end{equation}
  \begin{equation}
  \label{PEC_bc_3}
H_{30}^ -  = H_{30}^ + {e^{ - 2j{\beta _3}{h_3}}}
\end{equation}
\end{subequations}
By combining \eqref{BCforE},\eqref{BCforH} and \eqref{PEC_bc}, reflection coefficients can be derived as
\begin{subequations}
\label{Reflection coefficients}
  \begin{equation}
  \label{R_0}
R_0=\frac{H_{10}^-}{H_{10}^+}=\frac{2}{\xi_{10}}\frac{A_0}{B}+1
  \end{equation}
  \begin{equation}
  \label{R_n}
R_n=\frac{H_{1n}^-}{H_{10}^+}=\frac{2}{\xi_{1n}}\frac{A_n}{B}~~~(n\ne0)
\end{equation}
\end{subequations}
where
\begin{subequations}

  \begin{equation}
  \label{An}
\begin{array}{l}{A_n} = - M_{2-}^0 M_{2+}^n S_2 C_3 \xi_{20} - M_{2-}^0 M_{2+}^n S_2 S_3 \xi_{20}\xi_{30} \\ \times \sum\limits_m{\frac{M_{3+}^m M_{3-}^m}{\xi_{1m}}} - M_{3-}^0 M_{3+}^n S_3 C_2 \xi_{30} - M_{3-}^0 M_{3+}^n \\ \times S_2 S_3 \xi_{20}\xi_{30} \sum\limits_m {\frac{M_{2+}^m M_{2-}^m}{\xi_{1m}}} +M_{3-}^0 M_{2+}^n S_2 S_3 \xi_{20} \xi_{30} \\ \times \sum\limits_m{\frac{M_{2-}^m M_{3+}^m}{\xi_{1m}}} + M_{2-}^0 M_{3+}^n S_2 S_3 \xi_{20} \xi_{30} \sum\limits_m{\frac{M_{2+}^m M_{3-}^m}{\xi_{1m}}} \end{array}
\end{equation}

  \begin{equation}
  \label{Bn}
\begin{array}{l}{B} = S_2 C_3 \xi_{20} \sum\limits_m{\frac{M_{2+}^m M_{2-}^m}{\xi_{1m}}} + S_3 C_2 \xi_{30} \sum\limits_m{\frac{M_{3+}^m M_{3-}^m}{\xi_{1m}}} \\ + C_2 C_3 + S_2 S_3 \xi_{20} \xi_{30} (\sum\limits_m{\frac{M_{2+}^m M_{2-}^m}{\xi_{1m}}} \sum\limits_m{\frac{M_{3+}^m M_{3-}^m}{\xi_{1m}}} \\ - \sum\limits_m{\frac{M_{2+}^m M_{3-}^m}{\xi_{1m}}} \sum\limits_m{\frac{M_{2-}^m M_{3+}^m}{\xi_{1m}}})  \end{array}
\end{equation}
\end{subequations}

and
\begin{subequations}
\label{C and S}
  \begin{equation}
  \label{Si}
{S_i} = (1 - {e^{ - 2j{\beta _i}{h_i}}})~~~~(i=2,3)
\end{equation}

  \begin{equation}
  \label{Ci}
{C_i} = \frac{w_{i}}{P}(1 + {e^{ - 2j{\beta _i}{h_i}}}) ~~~~(i=2,3)
\end{equation}
\end{subequations}
Finally, The diffraction efficiencies (the ratio of diffracted power to the incident wave) can be calculated by the following relation:
\begin{equation}
DE_{n}=R_n R_n^* Re(\frac{k_{z1n}}{k_{z10}})  
\end{equation}

These calculations can be readily generalized to the case of multiple CMGs (a CMG that includes multi-slit in each unit cell) following a similar approach.

For TE polarization, like the TM polarization of incident wave, first, we define the electric and magnetic fields in each region, except that in regions II and III, we only take into account the first TE mode of the parallel plate waveguide. Therefore, the validity of the method is limited to the frequency range below the cutoff frequency of the mode inside the slits. This condition requires that $\lambda>2w_in_i$. Then, by applying boundary conditions, reflection coefficients can be derived \cite{hemmatyar2017phase}. For the sake of brevity, these calculations are not included here.

\subsection{Numerical Results}
In this section, we validate and verify the accuracy of the proposed circuit model through some numerical examples. We demonstrate that our proposed model produce very accurate results for $\lambda>2w_in_i$. 

\begin{figure} [hb!]

\centering\includegraphics[width=\linewidth]{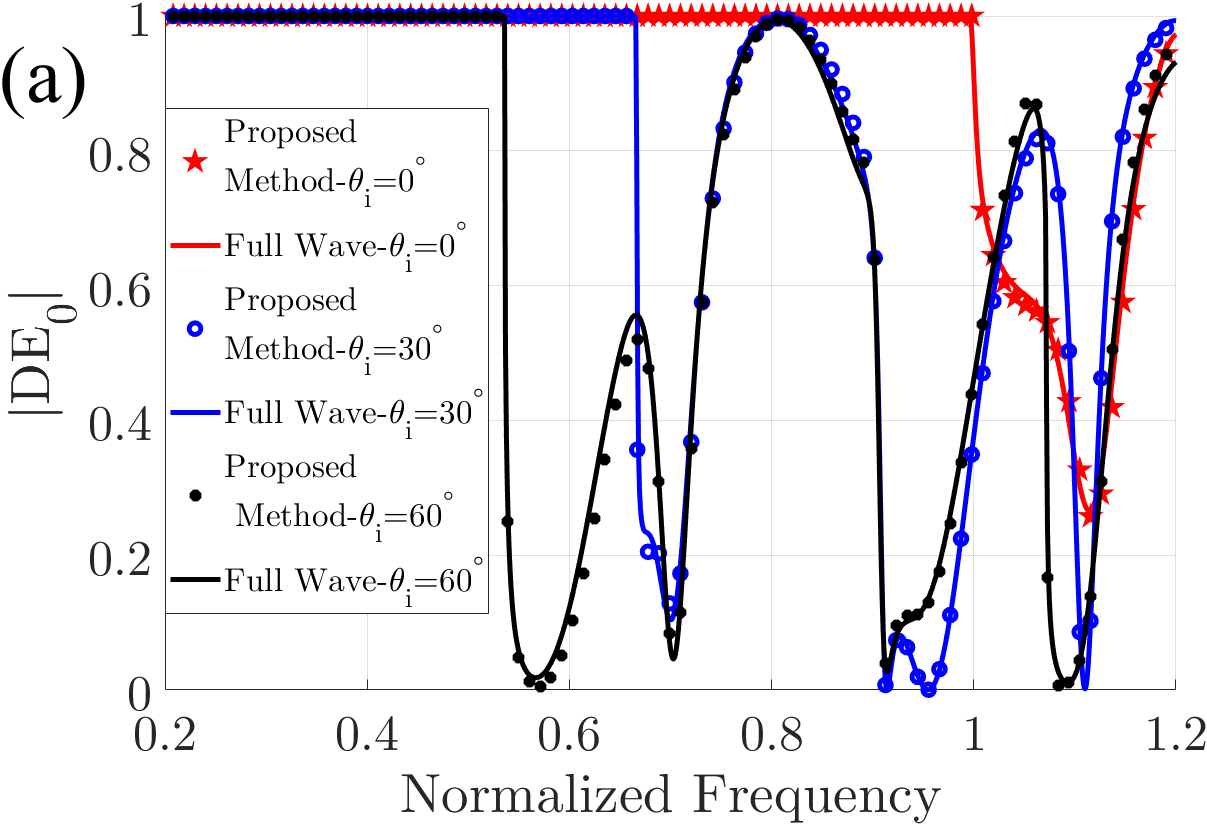}
\centering\includegraphics[width=\linewidth]{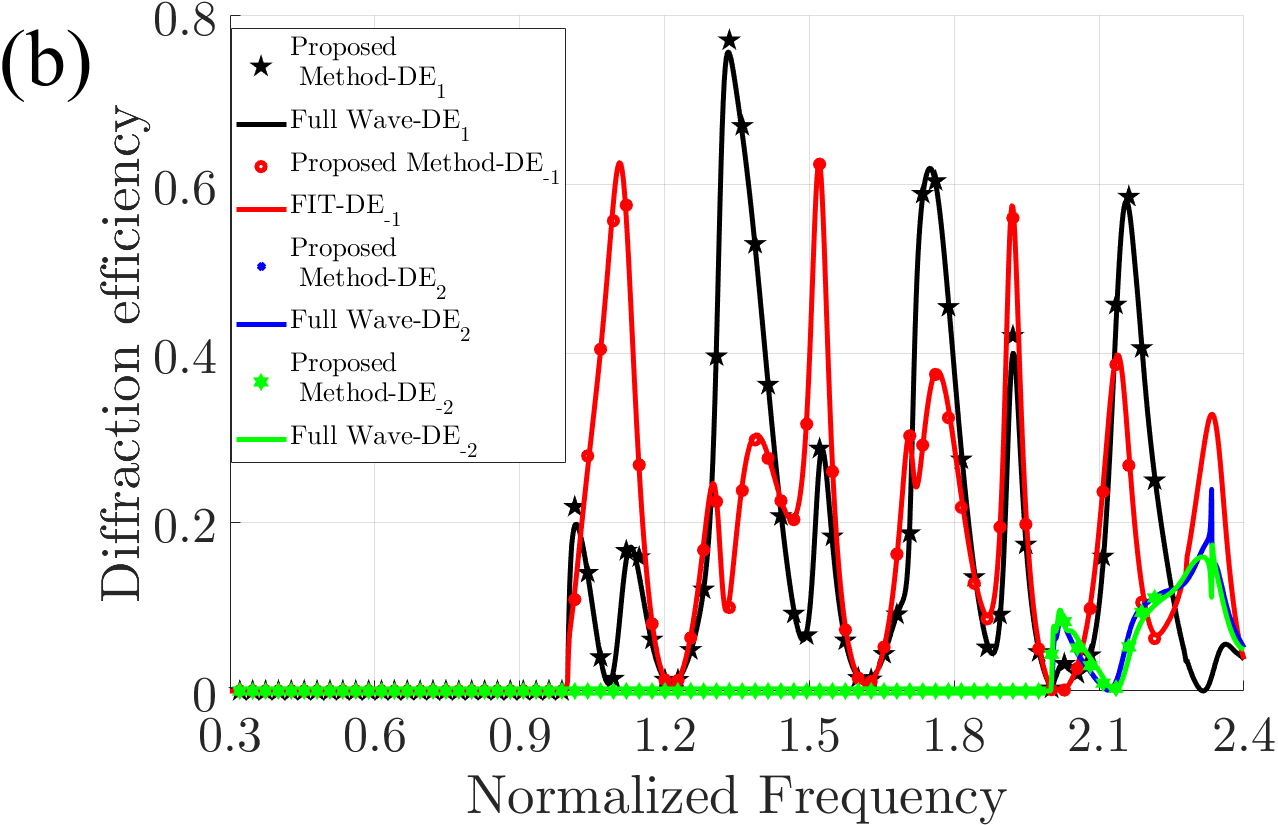}
\caption{Comparing the results of proposed method with full wave simulations: (a)The diffraction of zeroth-order mode for different incident angles: $\theta_i=0^\circ, 30^\circ $ and $~60^\circ$. (b)The diffraction efficiency of higher orders at normal incidence. The CMG parameters are assumed as $w_2=0.15P, w_3=0.2P, h_2=1.25P, h_3=1.625P, d=0.25P, n_1=n_2=1$ and $n_3=1.5$.}
\label{NE}
\end{figure}

Consider a CMG in accordance with Fig.\ref{CMG}. Here, the normalized frequency is defined as $\omega_n=P/\lambda$ and  the CMG parameters are set to $w_2=0.15P, w_3=0.2P$, $h_2=1.25P$, $h_3=1.625P$, $d=0.25P$, $n_1=n_2=1$ and $n_3=1.5$. The zeroth (specular) diffracted order obtained from the analytical method for different incident angles is plotted in Fig.{\ref{NE}}a. To validate the proposed analytical method in the previous subsections, a full-wave simulation is performed using the finite integration technique (FIT) in CST Microwave Studio 2019. Also, the first and the second diffraction efficiencies of the structure are plotted in Fig.\ref{NE}b for the normal incident. The excellent agreement that is observed between the FIT results and our analytical method, demonstrating the accuracy of the proposed model. The difference between $DE_1$ and $DE_{-1}$ can be observed from Fig\ref{NE}b. This difference stems from the geometrical asymmetry of the structure along $x$-direction, which can be highly useful in the design of metagrating and anomalous reflector.

\section{Design of perfect anomalous reflector}

In this section, we present an analytical method for the design of metagratings with anomalous reflection reaching to unity efficiency. Based on  FB theorem, a discrete set of modes can be diffracted when a plane wave impinges to a periodic structure. As shown in Fig.\ref{CMG}, except the specular mode (zero-order), all the FB modes diffract to a various angle while are not equal with the angle of the incident wave. If all the power of incident wave transfer to a propagating FB mode, the perfect anomalous reflection can be achieved. When the angle of the incident wave is 0, for a perfect anomalous reflection, the structure must have geometrical asymmetry in $x$-direction; otherwise, in the best situation, the efficiency of anomalous reflection is $50\%$ \cite{ra2017metagratings}.

The proposed CMG structure has geometrical asymmetry in $x$-direction and thus can be used in metagrating applications. For simplification of design process, we make two assumptions: 1- normal incidence, 2- $0$ (specular) and ($\pm1$) orders are the only propagating FB modes and higher-order modes are evanescent (at normal incidence this condition can be satisfied with $P<2\lambda$).  Our goal is to eliminate (0) and (1) modes and transfer all the power to $(-1)$ mode. First, we specify the period of the CMG based on the desired angle of anomalously reflected wave:
\begin{equation}
  P = \lambda/\sin(|\theta_{-1}|) 
  \label{Angle of (-1)}
\end{equation}
obtained from \eqref{kx1m}. For more simplicity in fabrication process, we assume that both slits are filled with air ($n_1=n_2=n_3=1$). As the main purpose, the (0) and (1) FB modes of the proposed structure should be zero. According to passivity condition and since the higher order modes(i.e., $n \geq 2$) are evanescent, by eliminating the $DE_0$ and $DE_{1}$, we achieve unitary efficieny for the $DE_{-1}$. Based on the theoretical formulation presented in the previous section, the genetic algorithm is utilized to minimize the $DE_0$ and $DE_{1}$ of the anomalous reflector. Hence, we define the cost function as $DE^2_{0}+DE^2_{1} $ in the desired frequency. The height and width of slits and also the distance between the centers of two slits are the variables in the optimization process.

\begin{figure} [ht!]

\centering\includegraphics[width=\linewidth]{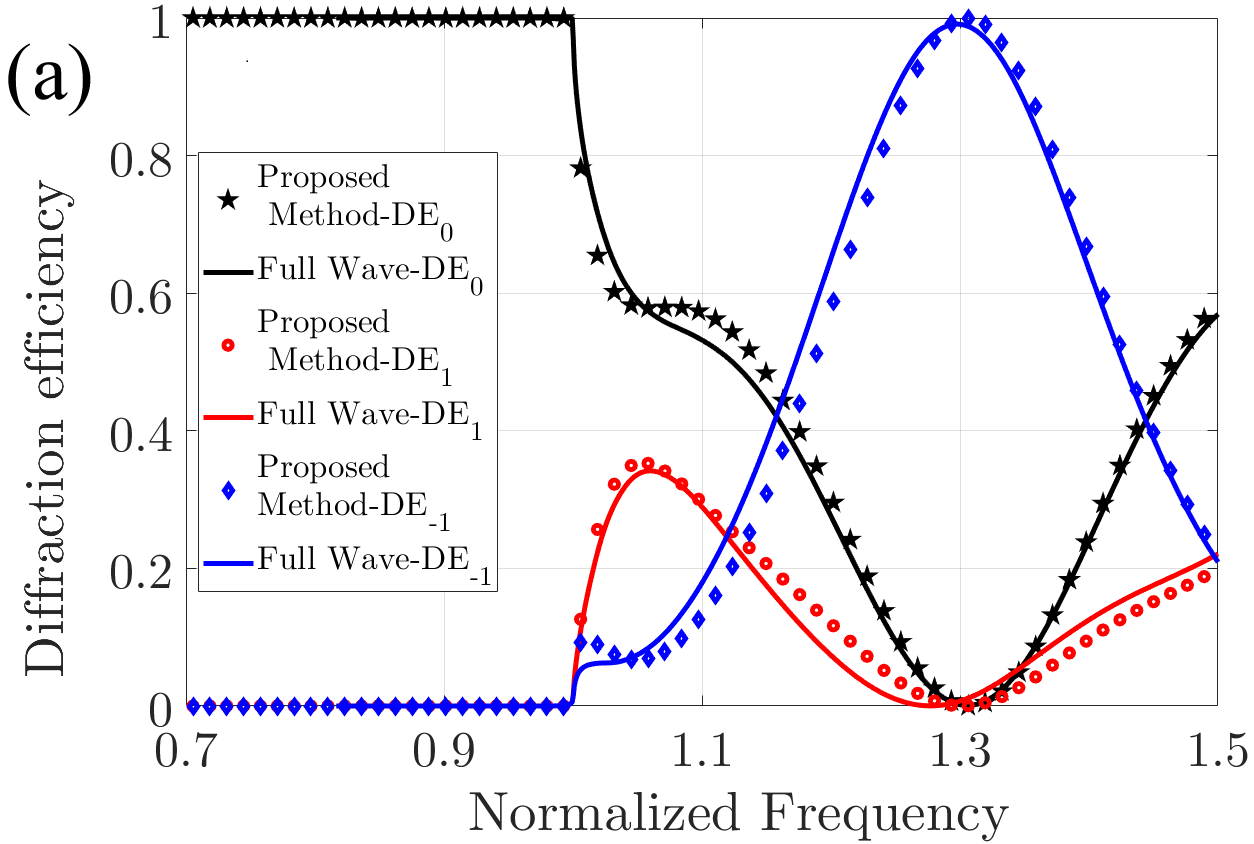}
\centering\includegraphics[width=\linewidth]{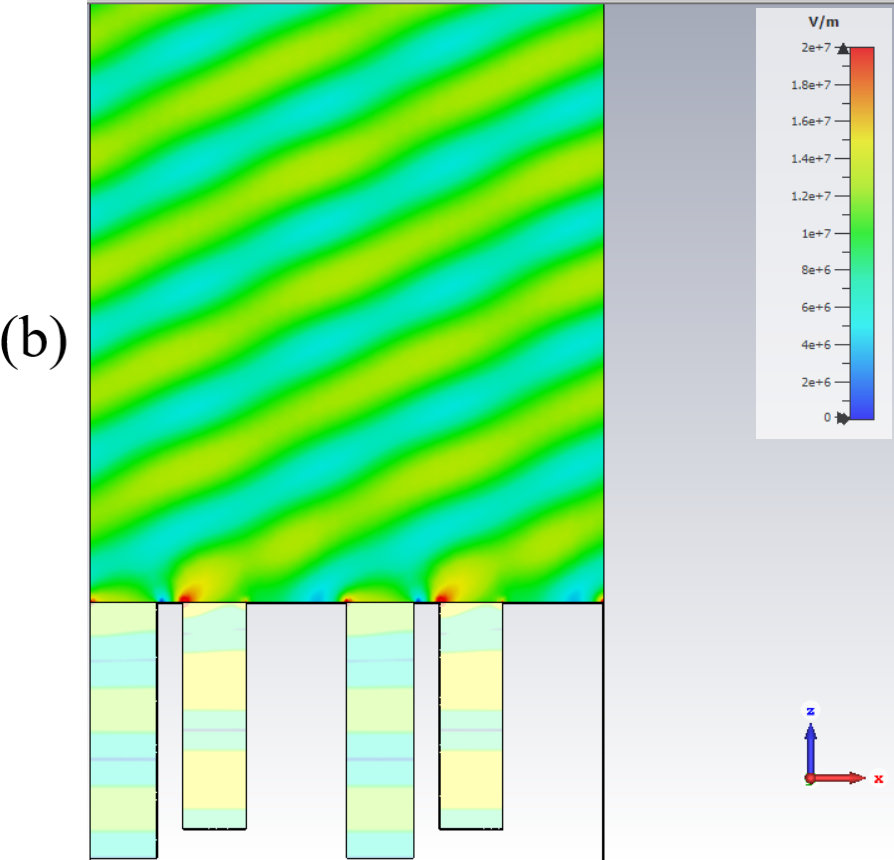}
\caption{(a) Diffraction efficiency of designed anomalous reflector. (b)The electrical field distribution in the $\omega_n=1.3$. The CMG parameters are designed as $w_2=0.26P, w_3=0.245P, h_2=0.99P, h_3=0.88P$ and $d=0.35P$ for transfer the incident power to the angle $-50^\circ$.}
\label{Perfect_AR}
\end{figure}

Using the above-mentioned design process we designed an anomalous reflector with near-to-unitary efficiency that reflects the incident wave to a direction with a $-50^\circ $ reflection angle. The parameters of  the designed CMG are $w_2=0.26P, w_3=0.245P, h_2=0.99P, h_3=0.88P$ and $d=0.35P$. The diffractions efficiency of designed CMG are shown in Fig.{\ref{Perfect_AR}}a. It can be seen from Fig.{\ref{Perfect_AR}}, that almost all the power ($99.91\%$) couples to the (-1) FB mode in the desired frequency $\omega_n=1.3$. This efficiency is a remarkable achievement in comparison with the previously published studies. The power efficiencies of anomalous reflection in \cite{chalabi2017efficient}, \cite{rabinovich2018analytical}, \cite{ra2018reconfigurable}, and \cite{chen2018polarization} are $94\%, 94.1\% ,94\%,$ and $74.8\%$, respectively. The distribution of reflected electric fields at the design frequency ($\omega_n=1.3$) is plotted in Fig.\ref{Perfect_AR}b which clearly shows anomalous reflection.

The CMG can be used for reflecting the normal incident to a direction with different angles. We extract the parameters of CMG similar to the previously designed anomalous reflector for several $\theta_{-1}$ values in the range $-40^\circ$ to $-80^\circ$. Also, CMG can anomalously reflect oblique incident electromagnetic waves. As a prototype, we design a metagrating that can reflect an incident wave from $\theta_i=50^\circ$ to $\theta_{-1}=-22.5^\circ$. The optimized structure parameters and the power efficiency of designed metagrating are listed in Table{\ref{table1}}. It can be observed from Table{\ref{table1}} that in all of the designed anomalous reflectors, we can achieve near to unitary efficiency. It should be noted that according to {\eqref{kx1m}}, the anomalous reflection occurs in the normalized frequency  $\omega_n=P/\lambda=1/|(sin\theta_i - sin\theta_{-1}|)$ for both of the oblique and the normal incident cases.

\begin{table} [ht!]
\caption{Optimum parameters for the anomalous reflection using CMG. For simplifying the fabrication process, the slits are filled with air.}

\setlength{\arrayrulewidth}{0.3mm}
	\setlength{\tabcolsep}{4pt}
	\renewcommand{\arraystretch}{1.75}

\begin{center}
\begin{tabular}{ c|c c c c c c c} 
 $\theta_i$ & $\theta_{-1}$ & $w_2/P$ & $w_3/P$ & $h_2/P$ & $h_3/P$ & $d/P$ & Power efficiency \\ \hline  
 $0^\circ$ & $-80^\circ$ & 0.17 & 0.28 & 0.81 & 1 & 0.42 & 99.99\% \\ 
  $0^\circ$ & $-70^\circ$ & 0.19 & 0.3 & 0.29 & 0.59 & 0.39 & 99.99\% \\ 
$0^\circ$ & $-60^\circ$ & 0.25 & 0.254 & 0.7 & 0.56 & 0.37 & 99.99\% \\ 
 $0^\circ$ & $-50^\circ$ & 0.26 & 0.245 & 0.99 & 0.88 & 0.35 & 99.91\% \\ 
$0^\circ$ & $-40^\circ$ & 0.12 & 0.09 & 0.143 & 0.1 & 0.35 & 99.9\% \\ 
 $50^\circ$ & $-22.5^\circ$ & 0.53 & 0.54 & 1 & 0.75 & 0.33 & 99.99\% \\ 
 \end{tabular}
\end{center}
\label{table1}
\end{table}

For reflecting the normal incident wave to the angle of below $30^\circ$, based on \eqref{Angle of (-1)}, it is required that $\lambda>2P$. According to \eqref{kx1m} and this condition, $\pm2$ orders of FB mode become propagating. Therefore, the design of this case is more challenging and thus there are little works on this case. In \cite{popov2018controlling}, it has been shown that a metagrating having $N$ polarization line currents per unit cell can eliminate the $N-1$ FB modes. However, this work did not propose a practical structure with specified parameters for the realization of anomalous reflection at angles below  $ 30^\circ$.

\begin{figure*} [ht!]  \includegraphics [width=15 cm]{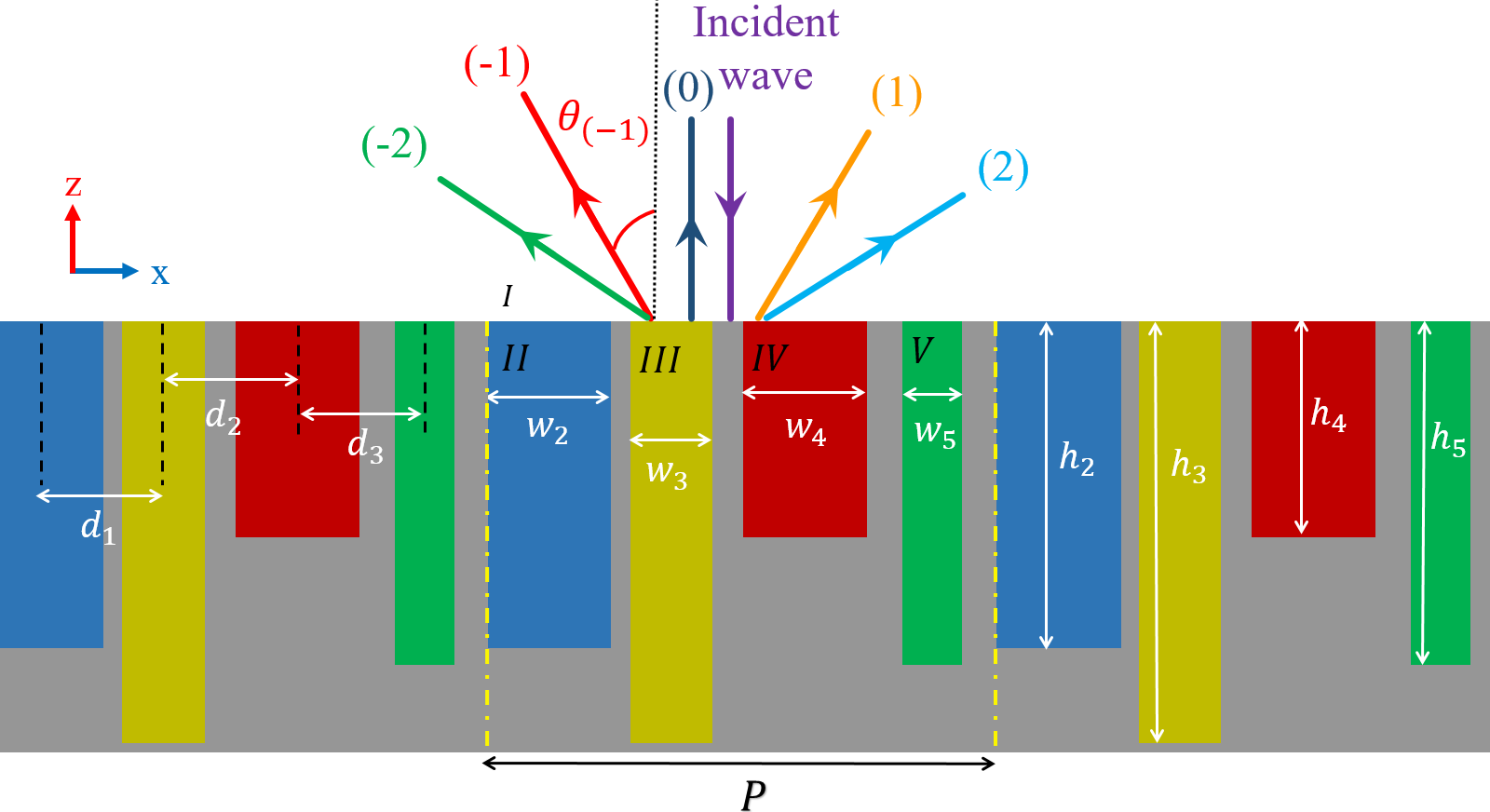}
\caption{The structure of multiple CMG with
four slits per period. CMG is illuminated by the normal incident.}
\label{Multi elemnt CMG}
\end{figure*}

For the design of an anomalous reflector at  $\theta_{-1}=-25^\circ$ for the normal incident waves, based on \eqref{Angle of (-1)}, periodicity must be chosen as $2.36\lambda$. It is noteworthy that similar to previous cases, our goal is to eliminate all FB mode except $ n=-1$.
The $(0), (\pm 1$) and ($\pm 2)$ FB modes are propagating for a structure with a period of $2.36\lambda$ . In this case, according to \cite{popov2018controlling}, the proposed structure of Fig.\ref{CMG} cannot be used. Hence, we propose a multi element compound metal grating as depicted in Fig.\ref{Multi elemnt CMG}. The proposed structure has four slits in a period. Again for easier fabrication, we assume the slits are filled with air.

\begin{figure} [ht!]

\centering\includegraphics[width=\linewidth]{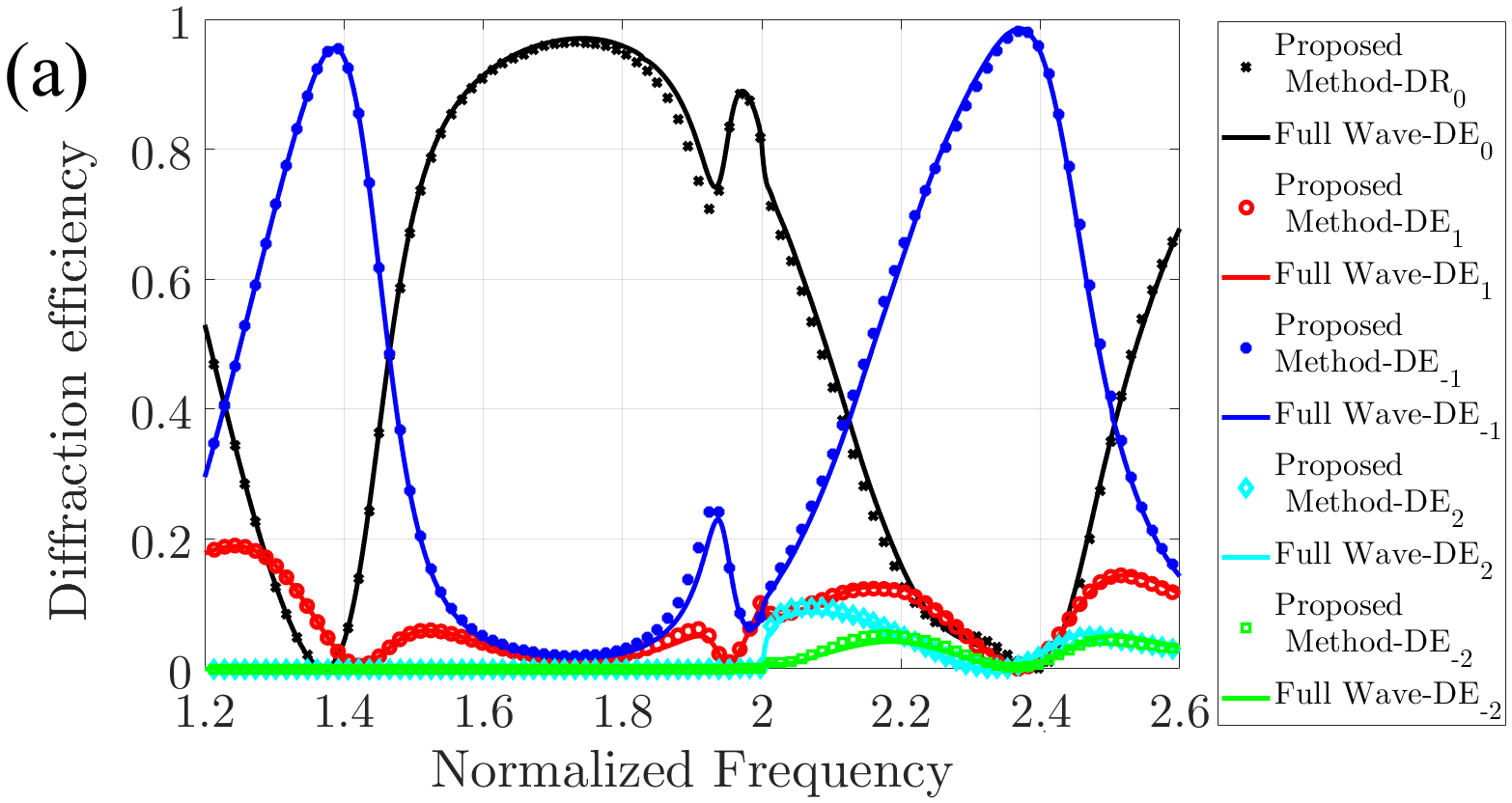}
\centering\includegraphics[width=\linewidth]{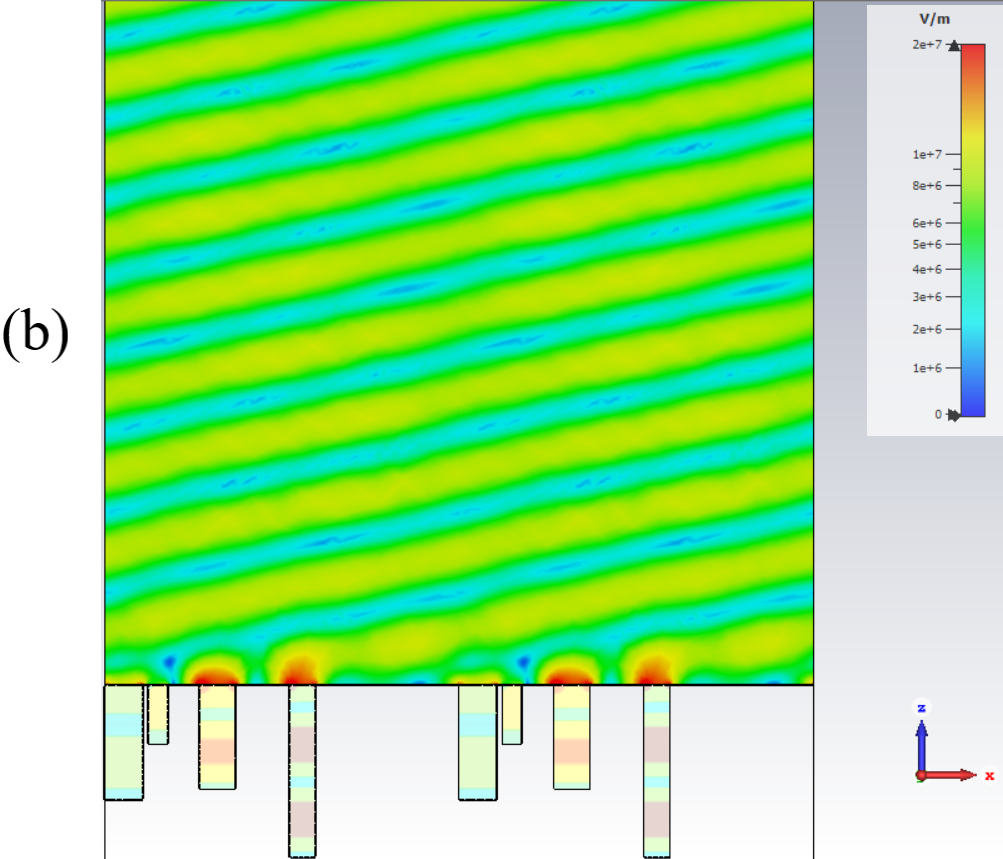}
\caption{(a)$DE_0,DE_{\pm 1}$ and $DE_{\pm 2}$ of the designed anomalous reflector for reflected angle of $-25^\circ$,  (b)The electrical field distribution at ($\omega_n=2.36$). The multiple CMG parameters are designed as $w_2=0.1P, w_3=0.05P, w_4=0.1P, w_5=0.07P, h_2=0.32P, h_3=0.16P,  h_4=0.29P, h_5=0.48P, d_1=0.1P, d_2=0.17P$ and $d_3=0.24P$}
\label{Perfect_AR2}
\end{figure}

The method of design is similar to the previously designed metagrating.
By eliminating $(0), (1),$ and $(\pm2)$ orders of the FB mode, the parameters of structure in Fig.\ref{Multi elemnt CMG} are designed as  $w_2=0.1P, w_3=0.05P, w_4=0.1P, w_5=0.07P, h_2=0.32P, h_3=0.16P,  h_4=0.29P, h_5=0.48P, d_1=0.1P, d_2=0.17P$ and $d_3=0.24P$ for reflecting incident wave into $-25^\circ$. The diffraction efficiencies of the propagating FB modes are plotted in Fig.\ref{Perfect_AR2}a. As can be observed from Fig.\ref{Perfect_AR2}a, the incident power goes to (-1) mode with excellent power efficiency ($98.55\%$). The electrical field distribution plotted in the design frequency demonstrates that the incident power is reflected at the angle of $-25^\circ$(Fig.\ref{Perfect_AR2}b).

\section{Conclusion}
In this paper, CMG was proposed for achieving perfect anomalous reflection as one of the most popular phenomena in the microwave, THz, and optic regime. An analytical method was proposed for analyzing CMG structures. The results of the proposed method were validated against full wave simulations, showing excellent agreement between the two solutions. Then, thanks to the proposed analytical method, perfect anomalous reflectors with virtually unitary power efficiency are presented to reflect the normal incident wave into the angles above $-30^\circ$. Moreover, multi-element CMG is proposed to reflect the normal incidence wave to angles below $-30^\circ$. Our analytical results demonstrated that perfect anomalous reflection could be realized using CMGs, which can  be very useful in the concepts of metagrating and wave manipulation.

\bibliographystyle{IEEEtran}
\bibliography{main,Bibliography}

\begin{thebibliography}{10}
\providecommand{\url}[1]{#1}
\csname url@samestyle\endcsname
\providecommand{\newblock}{\relax}
\providecommand{\bibinfo}[2]{#2}
\providecommand{\BIBentrySTDinterwordspacing}{\spaceskip=0pt\relax}
\providecommand{\BIBentryALTinterwordstretchfactor}{4}
\providecommand{\BIBentryALTinterwordspacing}{\spaceskip=\fontdimen2\font plus
\BIBentryALTinterwordstretchfactor\fontdimen3\font minus
  \fontdimen4\font\relax}
\providecommand{\BIBforeignlanguage}[2]{{%
\expandafter\ifx\csname l@#1\endcsname\relax
\typeout{** WARNING: IEEEtran.bst: No hyphenation pattern has been}%
\typeout{** loaded for the language `#1'. Using the pattern for}%
\typeout{** the default language instead.}%
\else
\language=\csname l@#1\endcsname
\fi
#2}}
\providecommand{\BIBdecl}{\relax}
\BIBdecl
\renewcommand{\BIBentryALTinterwordstretchfactor}{4}

\bibitem{rose2005wavelength}
B.~Rose, T.~Rasmussen, C.~Khalfaoui, M.~Rasmussen, J.~Bastue, P.~E. Ibsen, and
  C.~F. Pedersen, ``Wavelength division multiplexed device,'' Dec.~20 2005, uS
  Patent 6,978,062.

\bibitem{hawthorn2001littrow}
C.~Hawthorn, K.~Weber, and R.~Scholten, ``Littrow configuration tunable
  external cavity diode laser with fixed direction output beam,'' \emph{Review
  of scientific instruments}, vol.~72, no.~12, pp. 4477--4479, 2001.

\bibitem{rhee1994chirped}
J.-K. Rhee, T.~S. Sosnowski, T.~B. Norris, J.~A. Arns, and W.~S. Colburn,
  ``Chirped-pulse amplification of 85-fs pulses at 250 khz with third-order
  dispersion compensation by use of holographic transmission gratings,''
  \emph{Optics letters}, vol.~19, no.~19, pp. 1550--1552, 1994.

\bibitem{hum2007modeling}
S.~V. Hum, M.~Okoniewski, and R.~J. Davies, ``Modeling and design of
  electronically tunable reflectarrays,'' \emph{IEEE transactions on Antennas
  and Propagation}, vol.~55, no.~8, pp. 2200--2210, 2007.

\bibitem{li2015visible}
Z.~Li, E.~Palacios, S.~Butun, and K.~Aydin, ``Visible-frequency metasurfaces
  for broadband anomalous reflection and high-efficiency spectrum splitting,''
  \emph{Nano letters}, vol.~15, no.~3, pp. 1615--1621, 2015.

\bibitem{sun2012high}
S.~Sun, K.-Y. Yang, C.-M. Wang, T.-K. Juan, W.~T. Chen, C.~Y. Liao, Q.~He,
  S.~Xiao, W.-T. Kung, G.-Y. Guo \emph{et~al.}, ``High-efficiency broadband
  anomalous reflection by gradient meta-surfaces,'' \emph{Nano letters},
  vol.~12, no.~12, pp. 6223--6229, 2012.

\bibitem{wong2018perfect}
A.~M. Wong and G.~V. Eleftheriades, ``Perfect anomalous reflection with a
  bipartite huygens’ metasurface,'' \emph{Physical Review X}, vol.~8, no.~1,
  p. 011036, 2018.

\bibitem{li2018new}
X.~Li, M.~Memarian, and T.~Itoh, ``A new cavity resonance assisted by
  anisotropic metasurfaces,'' \emph{IEEE Transactions on Microwave Theory and
  Techniques}, vol.~66, no.~7, pp. 3224--3233, 2018.

\bibitem{mueller2017metasurface}
J.~B. Mueller, N.~A. Rubin, R.~C. Devlin, B.~Groever, and F.~Capasso,
  ``Metasurface polarization optics: independent phase control of arbitrary
  orthogonal states of polarization,'' \emph{Physical Review Letters}, vol.
  118, no.~11, p. 113901, 2017.

\bibitem{rahmanzadeh2018multilayer}
M.~Rahmanzadeh, H.~Rajabalipanah, and A.~Abdolali, ``Multilayer graphene-based
  metasurfaces: robust design method for extremely broadband, wide-angle, and
  polarization-insensitive terahertz absorbers,'' \emph{Applied optics},
  vol.~57, no.~4, pp. 959--968, 2018.

\bibitem{rahmanzadeh2018adopting}
M.~Rahmanzadeh, A.~Abdolali, A.~Khavasi, and H.~Rajabalipanah, ``Adopting image
  theorem for rigorous analysis of a perfect electric conductor--backed array
  of graphene ribbons,'' \emph{JOSA B}, vol.~35, no.~8, pp. 1836--1844, 2018.

\bibitem{momeni2019generalized}
A.~Momeni, H.~Rajabalipanah, A.~Abdolali, and K.~Achouri, ``Generalized optical
  signal processing based on multioperator metasurfaces synthesized by
  susceptibility tensors,'' \emph{Physical Review Applied}, vol.~11, no.~6, p.
  064042, 2019.

\bibitem{momeni2018information}
A.~Momeni, K.~Rouhi, H.~Rajabalipanah, and A.~Abdolali, ``An information
  theory-inspired strategy for design of re-programmable encrypted
  graphene-based coding metasurfaces at terahertz frequencies,''
  \emph{Scientific reports}, vol.~8, no.~1, p. 6200, 2018.

\bibitem{kildishev2013planar}
A.~V. Kildishev, A.~Boltasseva, and V.~M. Shalaev, ``Planar photonics with
  metasurfaces,'' \emph{Science}, vol. 339, no. 6125, p. 1232009, 2013.

\bibitem{sautter2015active}
J.~Sautter, I.~Staude, M.~Decker, E.~Rusak, D.~N. Neshev, I.~Brener, and Y.~S.
  Kivshar, ``Active tuning of all-dielectric metasurfaces,'' \emph{ACS nano},
  vol.~9, no.~4, pp. 4308--4315, 2015.

\bibitem{zheng2015metasurface}
G.~Zheng, H.~M{\"u}hlenbernd, M.~Kenney, G.~Li, T.~Zentgraf, and S.~Zhang,
  ``Metasurface holograms reaching 80\% efficiency,'' \emph{Nature
  nanotechnology}, vol.~10, no.~4, p. 308, 2015.

\bibitem{hosseininejad2019digital}
S.~E. Hosseininejad, K.~Rouhi, M.~Neshat, A.~Cabellos-Aparicio, S.~Abadal, and
  E.~Alarc{\'o}n, ``Digital metasurface based on graphene: An application to
  beam steering in terahertz plasmonic antennas,'' \emph{IEEE Transactions on
  Nanotechnology}, vol.~18, pp. 734--746, 2019.

\bibitem{sun2012gradient}
S.~Sun, Q.~He, S.~Xiao, Q.~Xu, X.~Li, and L.~Zhou, ``Gradient-index
  meta-surfaces as a bridge linking propagating waves and surface waves,''
  \emph{Nature materials}, vol.~11, no.~5, p. 426, 2012.

\bibitem{rouhi2019multi}
K.~Rouhi, H.~Rajabalipanah, and A.~Abdolali, ``Multi-bit graphene-based
  bias-encoded metasurfaces for real-time terahertz wavefront shaping: From
  controllable orbital angular momentum generation toward arbitrary beam
  tailoring,'' \emph{Carbon}, vol. 149, pp. 125--138, 2019.

\bibitem{li2015x}
H.~Li, G.~Wang, H.-X. Xu, T.~Cai, and J.~Liang, ``X-band phase-gradient
  metasurface for high-gain lens antenna application,'' \emph{IEEE Transactions
  on Antennas and Propagation}, vol.~63, no.~11, pp. 5144--5149, 2015.

\bibitem{yu2011light}
N.~Yu, P.~Genevet, M.~A. Kats, F.~Aieta, J.-P. Tetienne, F.~Capasso, and
  Z.~Gaburro, ``Light propagation with phase discontinuities: generalized laws
  of reflection and refraction,'' \emph{science}, vol. 334, no. 6054, pp.
  333--337, 2011.

\bibitem{asadchy2016perfect}
V.~S. Asadchy, M.~Albooyeh, S.~N. Tcvetkova, A.~D{\'\i}az-Rubio, Y.~Ra'di, and
  S.~Tretyakov, ``Perfect control of reflection and refraction using spatially
  dispersive metasurfaces,'' \emph{Physical Review B}, vol.~94, no.~7, p.
  075142, 2016.

\bibitem{epstein2016huygens}
A.~Epstein and G.~V. Eleftheriades, ``Huygens’ metasurfaces via the
  equivalence principle: design and applications,'' \emph{JOSA B}, vol.~33,
  no.~2, pp. A31--A50, 2016.

\bibitem{diaz2017generalized}
A.~D{\'\i}az-Rubio, V.~S. Asadchy, A.~Elsakka, and S.~A. Tretyakov, ``From the
  generalized reflection law to the realization of perfect anomalous
  reflectors,'' \emph{Science advances}, vol.~3, no.~8, p. e1602714, 2017.

\bibitem{ra2017metagratings}
Y.~Ra’di, D.~L. Sounas, and A.~Al{\`u}, ``Metagratings: Beyond the limits of
  graded metasurfaces for wave front control,'' \emph{Physical review letters},
  vol. 119, no.~6, p. 067404, 2017.

\bibitem{wood1902remarkable}
R.~W. Wood, ``On a remarkable case of uneven distribution of light in a
  diffraction grating spectrum,'' \emph{Proceedings of the Physical Society of
  London}, vol.~18, no.~1, p. 269, 1902.

\bibitem{khavasi2009adaptive}
A.~Khavasi and K.~Mehrany, ``Adaptive spatial resolution in fast, efficient,
  and stable analysis of metallic lamellar gratings at microwave frequencies,''
  \emph{IEEE transactions on antennas and propagation}, vol.~57, no.~4, pp.
  1115--1121, 2009.

\bibitem{harvey1991external}
K.~Harvey and C.~Myatt, ``External-cavity diode laser using a grazing-incidence
  diffraction grating,'' \emph{Optics letters}, vol.~16, no.~12, pp. 910--912,
  1991.

\bibitem{blanchard1999simultaneous}
P.~M. Blanchard and A.~H. Greenaway, ``Simultaneous multiplane imaging with a
  distorted diffraction grating,'' \emph{Applied optics}, vol.~38, no.~32, pp.
  6692--6699, 1999.

\bibitem{ra2018reconfigurable}
Y.~Ra’di and A.~Al{\`u}, ``Reconfigurable metagratings,'' \emph{ACS
  Photonics}, vol.~5, no.~5, pp. 1779--1785, 2018.

\bibitem{dong2019low}
X.~Dong, J.~Cheng, F.~Fan, and S.~Chang, ``Low-index second-order metagratings
  for large-angle anomalous reflection,'' \emph{Optics letters}, vol.~44,
  no.~4, pp. 939--942, 2019.

\bibitem{sell2018ultra}
D.~Sell, J.~Yang, E.~W. Wang, T.~Phan, S.~Doshay, and J.~A. Fan,
  ``Ultra-high-efficiency anomalous refraction with dielectric metasurfaces,''
  \emph{ACS Photonics}, vol.~5, no.~6, pp. 2402--2407, 2018.

\bibitem{neder2019combined}
A.~A. Neder, Y.~Ra’di and A.~Polman, ``Combined metagratings for efficient
  broad-angle scattering metasurface,'' \emph{ACS photonics}, vol.~6, no.~4,
  pp. 1010--1017, 2019.

\bibitem{inampudi2018neural}
S.~Inampudi and H.~Mosallaei, ``Neural network based design of metagratings,''
  \emph{Applied Physics Letters}, vol. 112, no.~24, p. 241102, 2018.

\bibitem{chalabi2017efficient}
H.~Chalabi, Y.~Ra'Di, D.~Sounas, and A.~Al{\`u}, ``Efficient anomalous
  reflection through near-field interactions in metasurfaces,'' \emph{Physical
  Review B}, vol.~96, no.~7, p. 075432, 2017.

\bibitem{rabinovich2018analytical}
O.~Rabinovich and A.~Epstein, ``Analytical design of printed circuit board
  (pcb) metagratings for perfect anomalous reflection,'' \emph{IEEE
  Transactions on Antennas and Propagation}, vol.~66, no.~8, pp. 4086--4095,
  2018.

\bibitem{rabinovich2019arbitrary}
------, ``Arbitrary diffraction engineering with multilayered multielement
  metagratings,'' \emph{arXiv preprint arXiv:1905.02376}, 2019.

\bibitem{rabinovich2019experimental}
O.~Rabinovich, I.~Kaplon, J.~Reis, and A.~Epstein, ``Experimental demonstration
  and in-depth investigation of analytically designed anomalous reflection
  metagratings,'' \emph{Physical Review B}, vol.~99, no.~12, p. 125101, 2019.

\bibitem{chen2018polarization}
J.~Chen, Y.~Zhang, Y.~Wang, F.~Kong, Y.~Jin, P.~Chen, J.~Xu, S.~Sun, and
  J.~Shao, ``Polarization-independent two-dimensional diffraction
  metal-dielectric grating,'' \emph{Applied Physics Letters}, vol. 113, no.~4,
  p. 041905, 2018.

\bibitem{popov2018controlling}
V.~Popov, F.~Boust, and S.~N. Burokur, ``Controlling diffraction patterns with
  metagratings,'' \emph{Physical Review Applied}, vol.~10, no.~1, p. 011002,
  2018.

\bibitem{molero2016dynamical}
C.~Molero, R.~Rodr{\'\i}guez-Berral, F.~Mesa, and F.~Medina, ``Dynamical
  equivalent circuit for 1-d periodic compound gratings,'' \emph{IEEE
  Transactions on Microwave Theory and Techniques}, vol.~64, no.~4, pp.
  1195--1208, 2016.

\bibitem{yarmoghaddam2014circuit}
E.~Yarmoghaddam, G.~K. Shirmanesh, A.~Khavasi, and K.~Mehrany, ``Circuit model
  for periodic array of slits with multiple propagating diffracted orders,''
  \emph{IEEE Transactions on Antennas and Propagation}, vol.~62, no.~8, pp.
  4041--4048, 2014.

\bibitem{hemmatyar2017phase}
O.~Hemmatyar, B.~Rahmani, A.~Bagheri, and A.~Khavasi, ``Phase resonance tuning
  and multi-band absorption via graphene-covered compound metallic gratings,''
  \emph{IEEE Journal of Quantum Electronics}, vol.~53, no.~5, pp. 1--10, 2017.

\end{thebibliography}

\end{document}